\documentclass[conference]{IEEEtran}
\usepackage{times}

\usepackage[numbers]{natbib}
\usepackage{multicol}
\usepackage[bookmarks=true]{hyperref}
\usepackage{graphicx}
\usepackage{amsmath}
\usepackage{xcolor} 
\usepackage{macros}
\usepackage{amsfonts}
\usepackage{algorithm, algpseudocode}
\usepackage{multirow}
\usepackage{arydshln}

\usepackage{amsmath}
\usepackage{amssymb}
\usepackage{hyperref}

\usepackage{subcaption}

\begin{document}

\title{Evader-Agnostic Team-Based Pursuit Strategies in Partially-Observable Environments}






\author{\authorblockN{Addison Kalanther,
Daniel Bostwick,
Chinmay Maheshwari, and
Shankar Sastry
\authorblockA{University of California Berkeley\\Emails: \{addikala@, daniel.k.bostwick@, chinmay\_maheshwari@, sastry@coe.\}berkeley.edu}}}

\maketitle
\renewcommand{\thefootnote}{\fnsymbol{footnote}}
\footnotetext[1]{This research is supported by the DARPA project  ANSR under grant number FA 8750-23-C-0080 administered by Air Force Research Laboratory, with project title "Provably Correct Design of Adaptive Hybrid Neuro-Symbolic Cyber Physical Systems".
}
\renewcommand{\thefootnote}{\arabic{footnote}}

\begin{abstract}
We consider a scenario where a team of two unmanned aerial vehicles (UAVs) pursue an evader UAV within an urban environment. Each agent has a limited view of their environment where buildings can occlude their field-of-view. Additionally, the pursuer team is agnostic about the evader in terms of its initial and final location, and the behavior of the evader. Consequently, the team needs to gather information by searching the environment and then track it to eventually intercept. To solve this multi-player, partially-observable, pursuit-evasion game, we develop a two-phase neuro-symbolic algorithm centered around the principle of bounded rationality. First, we devise an offline approach using deep reinforcement learning to progressively train adversarial policies for the pursuer team against fictitious evaders. This creates $k$-levels of rationality for each agent in preparation for the online phase. Then, we employ an online classification algorithm to determine a "best guess" of our current opponent from the set of iteratively-trained strategic agents and apply the best player response. Using this schema, we improved average performance when facing a random evader in our environment.

{\color{blue} Website: }\url{https://sastry-group.github.io/PEG/}
\end{abstract}

\section{Introduction}
Pursuit-evasion games serve as fundamental models for a wide range of real-world scenarios, from security and surveillance to search and rescue. In these games, a set of 
pursuers attempts to capture one or more evaders, who in turn aim to avoid capture by exploiting the limitations of the pursuers. 
Urban environments pose particularly challenging settings for such interactions due to different kinds of occlusions that could be accessible/inaccessible or viewable/non-viewable by pursuers and evaders as seen in \cite{yan2023long}. 
Most studies have either (i) considered full/limited information about evaders at all times \cite{9303044, li2021cfr, xue2021solving, zhang2017optimal, zhang2019optimal, li2024grasper, li2023solving} (ii) focused on simple scenarios with simple or no obstacles \cite{li2020uav, quintero2015robust, kokolakis2020bounded} (iii) focused on the problem of computing one fixed policy (such as  (approximate) Nash equilibrium) and does not adapt its policy, \cite{9303044, li2021cfr, xue2021solving, zhang2017optimal, zhang2019optimal, li2024grasper, li2023solving} (iv) considered a pursuer team comprised of homogeneous agents \cite{lopez2019solutions, li2021cfr, xue2021solving, zhang2019optimal, li2024grasper}. But, often times, these assumptions are too restrictive to deploy these technologies in the real world. Indeed, the pursuers would need to gather information about the evader without any initial knowledge, work in complicated urban environments with limited accessibility and viewability, adapt their strategy in real-time using new information, and have heterogeneous roles for the team members.  

\section{Problem Formulation}


We consider a heterogeneous pursuer team consisting of two drones operating
at different altitudes: a High-Level Pursuer (HLP) and a Low-Level Pursuer (LLP).
The HLP operates at a higher altitude than the LLP and the Evader, providing it
with a larger field-of-view (FOV) and immunity from obstacles. In contrast, the
LLP flies at a lower altitude (same as that of Evader), which leads to smaller field-of-view
than the HLP, but this allows it to intercept the Evader.

The HLP is equipped with a downward-facing camera and flies above the map,
with the role of detecting and tracking the Evader from above. The LLP
and Evader are equipped with a forward-facing camera. Upon seeing the Evader,
the HLP relays this info to the LLP. The two pursuers maintain communication
to share their own state and any available information about the Evader’s
location and movement.

\begin{figure}[htbp]
    \centering
    \includegraphics[scale=0.2]{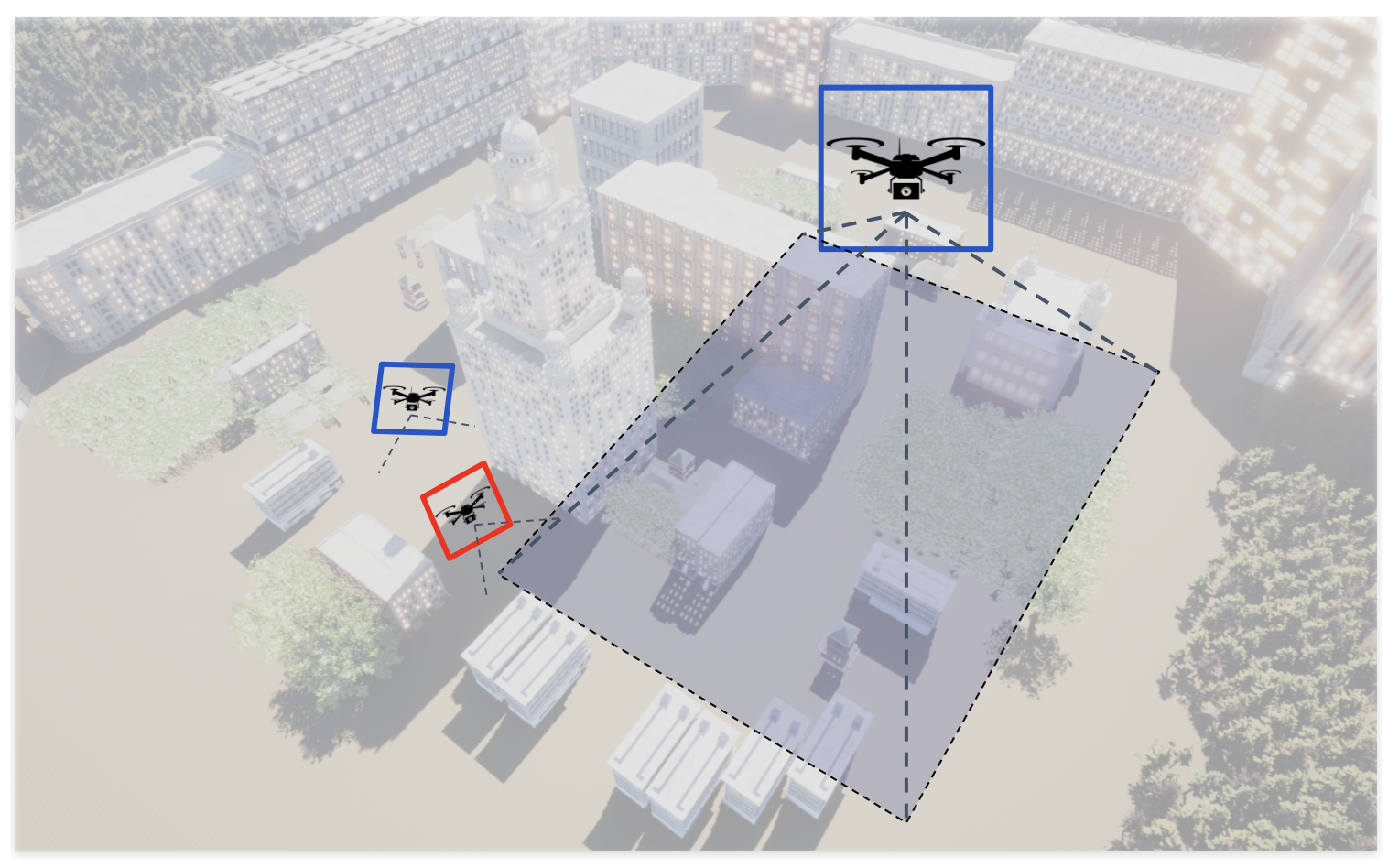}
    \caption{Visualization of scenario inside AirSim simulator.}
    \label{fig:airsim}
\end{figure}

The urban environment contains heterogeneous obstacles that impose access and visibility
constraints on the pursuer and Evader drones. These constraints necessitate coordination between the HLP and LLP, as only one might be able to view the Evader. This environment is depicted in Fig.~\ref{fig:airsim}, where the HLP is seen to have an advantage in uncovered areas, but the LLP can see the Evader when under foliage such as trees. 

We created a grid-based simulator which constains obstacles with similar rules pertaining to accessibility and viewability. Here, we deployed policies $\policyHLP, \policyLLP, \policyEDR$ for the HLP, LLP, and Evader respectively. The observations $\textbf{o}(t)$ for each agent only include what is seen directly in their FOV each timestep $t$, making the game partially-observable. For this reason, the policies input histories $\textbf{h}(t)$ containing observations $\{\textbf{o}(t'):t' \in [0, \dots, t]\}$. Each team is permitted to share information between their agents, so the LLP (resp. HLP) has access to observations $\obsHLP$ (resp. \(\obsLLP\)).

The goal of this paper is to learn pursuer policies $(\piHLP, \piLLP)$ capable of intercepting an unknown evader $\piEDR$. Given the environment, this presents several challenges.

\textbf{1. Modeling evader behavior}. As $\piEDR$ is not known, pursuers must be trained against a sufficiently representative set of potential evaders.

 \textbf{2. Unknown evader start and goal locations.} Trained policies of pursuers must be robust to
    arbitrary origin-destination locations of the Evader.

\textbf{3. Long-term planning under partial observability.} Pursuers must operate in a partially-observable environment, where information gathering is key to catching the evader.

\textbf{4. Asymmetric pursuer agents.} The HLP and LLP have differing roles and capabilities. Even under shared policy frameworks, this presents significant challenges.

\section{Approach}
To tackle these challenges, we propose a two-phased approach comprising of offline and online phases. 

\paragraph{Offline Phase}
The offline phase is based on: (i) a well-known behavioral game theory concept of level-$k$ reasoning, and (ii) a classifier module to predict the level of which the evader is closest to.

The level-$k$ reasoning models agents by assigning each agent a reasoning level $k
\in \mathbb{N}$ \cite{simon1997models}. A level-0 agent follows a fixed, non-strategic policy—e.g., uniformly
random actions or a predefined heuristic. Additionally, a level-$k$ agent assumes that all other agents
operate at level-$(k{-}1)$, and computes a policy that optimizes under this assumption.
This creates a hierarchy of sophisticated policies, each of which can be trained using a POMDP framework in RL. However, due to the asymmetry in our teams, we slightly abuse this notation for convenience. We say $(\piHLP^{(i)}, \piLLP^{(i)})$ is a response to $\piEDR^{(i)}$ while $\piEDR^{(i)}$ is a response to $(\piHLP^{(i - 1)}, \piLLP^{(i - 1)})$, which follows the convention above. Details about this and the training can be found in our extended, upcoming paper \cite{peg}.


In order to deploy a best response policy against an unknown opponent in the online phase, we train a classifier
$\mathcal{C}_{t}: (\histHLP(t), \histLLP(t)) \rightarrow \{0, 1, ..., K\}$ to
take in the observation histories of the HLP and LLP up to time $t$ and output the
level that the currently faced opponent is most similar to.

\paragraph{Online Phase}
For the online phase, we aim to classify our current opponent as one of the
evader policies we have seen during offline training, and then deploy our best response
policy against the selected evader, which was generated via reinforcement
learning. This is done by using the classifier $\mathcal{C}_{t}$ to predict the currently
seen evader $\policyEDR^{(i)}$ as being one of $\policyEDR^{(0)}$ to $\policyEDR^{(K)}$,
and deploying $(\policyLLP^{(i)}, \policyHLP^{(i)})$ against it. 

\section{Analysis}

We test this regime up to two levels of agents, meaning two policies for the evader and two responding policies for the pursuer team. We first analyze the efficacy of the offline phase independently. To determine performance, we evaluated based on some key metrics such as, pursuer winrate, percentage of episodes where the pursuers are able to see the Evader, time at which the Evader was seen first, percentage of time when the Evader was in the FOV.

Given these metrics, we summarize the performance in Table I. As expected, the trained responses performed well against that of which it is trained. However, pursuer performance degenerates whenever it faces an opponent on which it is not trained on. This motivates the need for our online classification module, which aims to serve as a meta-algorithm to determine the best policy to deploy against an unknown evader based on the history of observations.

\begin{table}[ht]
\centering
\caption*{\textbf{Pursuer vs. Evader Policy Performance Summary}} 
\vspace{0.5em}
\resizebox{\linewidth}{!}{
\begin{tabular}{ |c|c|c|c|c|c| } 
\hline
\textbf{Scenario} & \textbf{Phase} & \textbf{Pursuer Win} & \textbf{Evader Seen} & \textbf{First Seen} & \textbf{Time in FOV} \\
\hline
\multirow{2}{10em}{$(\policyHLP^{(0)}, \policyLLP^{(0)})$ vs. $\policyEDR^{(0)}$}
& Offline & \textbf{80\%} & \textbf{88\%} & \textbf{3.45} & \textbf{45\%} \\
& Online & \textbf{80\%} & \textbf{88\%} & 3.94 & 44\% \\
\hline
\multirow{2}{10em}{$(\policyHLP^{(0)}, \policyLLP^{(0)})$ vs. $\policyEDR^{(1)}$} 
& Offline & 3\% & 68\% & \textbf{4.53} & 7.6\% \\ 
& Online & \textbf{28\%} & \textbf{75\%} & 6.67 & \textbf{14\%} \\
\hline
\multirow{2}{10em}{$(\policyHLP^{(1)}, \policyLLP^{(1)})$ vs. $\policyEDR^{(1)}$}
& Offline & \textbf{60\%} & \textbf{83\%} & \textbf{4.78} & \textbf{41\%} \\ 
& Online & 45\% & \textbf{83\%} & 5.82 & 19\% \\
\hline
\multirow{2}{10em}{$(\policyHLP^{(1)}, \policyLLP^{(1)})$ vs. $\policyEDR^{(0)}$} 
& Offline & 10\% & \textbf{58\%} & \textbf{3.3} & 15\% \\ 
& Online & \textbf{18\%} & 53\% & 5.71 & \textbf{25\%} \\
\hline
\end{tabular}
}
\vspace{0.5em}
\caption{Scenario $(\policyHLP^{(i)}, \policyLLP^{(i)})$ vs. $\policyEDR^{(j)}$ references policies for pursuer level $i$ and evader level $j$ respectively. Bold highlights best figure between the offline the online phase within the same scenario.}
\label{tab:pursuer_evader_metrics}
\end{table}

\begin{figure}[htbp]
    \centering
    \includegraphics[scale=0.35]{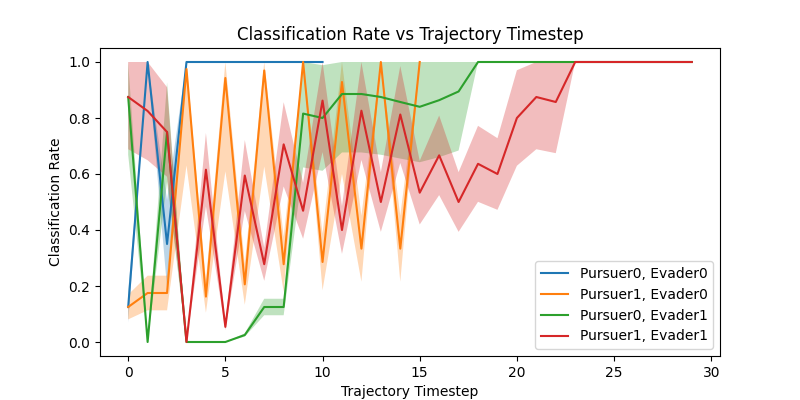}
    \caption{Classification rate vs Trajectory Timestep.}
    \label{fig:online_classification}
\end{figure}

The online classification module is trained by randomly choosing a pursuer level and an evader level and collecting observation histories from games with these players. We then train a classification model to take these histories and output the level of evader it was facing. In training, we see a classification rate of 98\%. However, in the online setting, it takes some time for the classifier to correctly classify the evader type, as seen in Fig.~\ref{fig:online_classification}. We attribute this to the fact that the classification module was trained with a dataset where the pursuers do not switch policies in the middle of the episode, which occurs in the online implementation. Ongoing efforts include extending this module to correctly classify an evader in the presence of policy switching to improve performance. 

However, despite these mixed results, we do see distinct improvement. While performance dropped for the cases where the pursuer starts with the best response for the evader, it significantly improved in the scenarios where the pursuer is not the best response. This suggests that, if the online classification module better covered the distribution of histories with switching policies, the performance of this schema can be greatly increased, which is an ongoing effort.




\bibliographystyle{plainnat}
\bibliography{references}

\end{document}